\documentclass[namedreferences]{solarphysics}
\usepackage[optionalrh]{spr-sola-addons} 
\usepackage{epsfig}          
\usepackage{graphicx}        
\usepackage{natbib}         
\usepackage{amssymb}        
\usepackage{color}           
\usepackage{url}             
\usepackage[T1]{fontenc}        

\newcommand{\mgxii}{{Mg~{\sc xii}}}

\newcommand{\sixii}{{Si~{\sc xii}}}
\newcommand{\sixiii}{{Si~{\sc xiii}}}
\newcommand{\sixiv}{{Si~{\sc xiv}}}
\newcommand{\sxiv}{{S~{\sc xiv}}}
\newcommand{\sxv}{{S~{\sc xv}}}

\newcommand{\arxvii}{{Ar~{\sc xvii}}}
\newcommand{\caxx}{{Ca~{\sc xx}}}
\newcommand{\caxix}{{Ca~{\sc xix}}}
\newcommand{\caxviii}{{Ca~{\sc xviii}}}
\newcommand{\caxvii}{{Ca~{\sc xvii}}}
\newcommand{\caxvi}{{Ca~{\sc xvi}}}
\newcommand{\fexxii}{{Fe~{\sc xxii}}}

\newcommand{\fexxiv}{{Fe~{\sc xxiv}}}
\newcommand{\fexxv}{{Fe~{\sc xxv}}}

\begin{document}

\begin{article}

\begin{opening}

\title{\bf X-ray Flare Spectra from the DIOGENESS Spectrometer and its concept applied to ChemiX on the Interhelioprobe spacecraft}

\author{Janusz~\surname{Sylwester}$^{1}$\sep Zbigniew~\surname{Kordylewski}$^{1}$\sep Stefan~\surname{P{\l}ocieniak}$^{1}$\sep Marek~\surname{Siarkowski}$^{1}$\sep Miros{\l}aw~\surname{Kowali\'{n}ski}$^{1}$\sep Stanis{\l}aw~\surname{Nowak}$^{1}$\sep Witold~\surname{Trzebi\'{n}ski}$^{1}$\sep  Marek~\surname{\'{S}t\k{e}\'{s}licki}$^{1}$\sep
Barbara~\surname{Sylwester}$^{1}$\sep Eugeniusz~\surname{Sta\'{n}czyk}$^{1,5}$\sep Ryszard~\surname{Zawerbny}$^{1,5}$\sep \.{Z}aneta~\surname{Szaforz}$^{1}$\sep Kenneth J. H.~\surname{Phillips}$^{2}$\sep Frantisek~\surname{Farnik}$^{3}$\sep  Anatolyi~\surname{Stepanov}$^{4}$\sep}

\runningauthor{J. Sylwester {\it et al.}}
\runningtitle{DIOGENESS crystal spectrometer and future spacecraft}

\institute{$^{1}$ Space Research Centre, Polish Academy of Sciences, Kopernika 11, 51-622 Wroc{\l}aw, Poland
                     email: \url{js@cbk.pan.wroc.pl}
            $^{2}$ Earth Sciences Department, Natural History Museum, London SW7 5BD, UK
                     email: \url{kennethjhphillips@yahoo.com}
            $^3$ Astronomical Institute, Czech Academy of Sciences, Ondrejov, Czech Republic
                    email: \url{ff@ai.cz}
             $^4$ Institute of Terrestrial Magnetism and Radiowave Propagation (IZMIRAN), Troisk, Moscow, Russia
                    email: \url{as@izmiran.ru}
             $^5$ Deceased}



\begin{abstract}
The {\em DIOGENESS} X-ray crystal spectrometer on the {\em CORONAS-F} spacecraft operated for a single month (25~August to 17~September) in 2001 but in its short lifetime obtained one hundred and forty high-resolution spectra from some eight solar flares with {\em GOES} importance ranging from C9 to X5. The instrument included four scanning flat crystals with wavelength ranges covering the regions of \sixiii\ (6.65~\AA), \sxv\ (5.04~\AA), and \caxix\ (3.18~\AA) X-ray lines and associated dielectronic satellites. Two crystals covering the \caxix\ lines were  oriented in a ``Dopplerometer'' manner, i.e. such that spatial and spectral displacements both of which commonly occur in flares can be separated. We describe the {\em DIOGENESS} spectrometer and the spectra obtained during flares which include lines not hitherto seen from spacecraft instruments. An instrument with very similar concept is presently being built for the two Russian  {\em Interhelioprobe} spacecraft due for launch in 2020 and 2022 that will make a near-encounter (perihelion $\sim 0.3$ a.u.) to the Sun in its orbit. We outline the results that are likely to be obtained.
\end{abstract}
\end{opening}

\section{Introduction}

Scanning X-ray flat crystal spectrometers on rockets and spacecraft have been used to observe flares and non-flaring active regions for many years, and the resulting spectra have helped our understanding of the energetics and physical characteristics of the emitting regions and the mechanism of flare energy release. Instruments with an accurate absolute intensity calibration have also enabled element abundances in the corona to be found. The rapidly developing initial phases of flares, in which high-temperature plasma is ejected from localised sources such as the footpoints of magnetic flux tubes, requires X-ray spectrometers that can acquire and store data on very short time scales. This has led to the application of bent crystal spectrometers with position-sensitive detectors for solar-dedicated spacecraft. Early examples include those on the Solar Maximum Mission (BCS or Bent Crystal Spectrometer: \inlinecite{act80}), {\em Yohkoh} (BCS or Bragg Crystal Spectrometer: \inlinecite{cul91}), and {\em CORONAS-F} ({\em RESIK} or REntgenovsky Spektrometr s Izognutymi Kristalami: \inlinecite{jsyl05}).

Bent crystal spectrometers have the advantage of obtaining spectra over their wavelength ranges instantaneously in a particular data-gathering interval, typically a few seconds, and also of generally having higher sensitivity than flat crystal spectrometers. However, for observing the initial phases of flares, spectra from bent crystal spectrometers, especially those without a fine collimator which reduces their sensitivity, are subject to a confusion of spectral and spatial blurring without the possibility of untangling the two unless independent measurements are available. This difficulty may be overcome with the simultaneous observation of flares with a pair of crystals oriented in such a way that the spatial displacements occur to longer wavelengths for one of them and to shorter wavelengths for the other. This ``dopplerometer'' mode was proposed for an instrument initially launched on a Vertical rocket in the Russian space programme by the Space Research Centre of the Polish Academy of Sciences in 1981. A later version, called {\em DIOGENESS} (DIagnostics Of Global ENErgy Sources and Sinks) was built by SRC for inclusion on the Russian {\em CORONAS-I} spacecraft in the mid-1990s and, following the failure of the instrument, for the subsequent {\em CORONAS-F} spacecraft, which was operational from 2001 till 2006. Four scanning flat crystals were included for the observation of the vicinity of intense lines of high-temperature ions -- \sixiii\ (6.65~\AA), \sxv\ (5.04~\AA), and \caxix\ (3.18~\AA) -- prominent in flare and some active region spectra, with two of the crystals observing the \caxix\ lines in a dopplerometer arrangement.

The {\em CORONAS-F} spacecraft was launched on 31 July 2001 and worked successfully until the end of 2006, with {\em RESIK} and {\em DIOGENESS} included in the instrument package. While {\em RESIK} obtained spectra until May 2003, {\em DIOGENESS} operated for only a few weeks as a fault in the scanning drive mechanism occurred on 17~September 2001. However, eight flares with {\em GOES} importance up to X5.5 were observed and one hundred and forty spectra were obtained in four wavelength channels. Detailed analysis was delayed for some years while the more extensive {\em RESIK} data set were examined. In the meantime, another version of {\em DIOGENESS} called {\em ChemiX} is being built for inclusion in the instrument package on two Russian {\em Interhelioprobe} spacecraft, due for launch in 2020 and 2022, which will eventually orbit the Sun in highly elliptical paths, with $30^{\circ}$ inclination to the ecliptic plane, bringing them to within 0.3 astronomical unit from the Sun. It is appropriate, therefore, to describe the {\em DIOGENESS} and {\em ChemiX} instruments in some detail at this time, even though the two earlier versions were launched some years ago. In particular it is of interest to describe the {\em DIOGENESS} spectra which have particularly good resolution and allow the flare dynamics and abundance ratios to be determined as well as investigation of lines not previously noted in solar spectra.

In this paper, we give a description of the {\em DIOGENESS} instrument and some details of the {\em CORONAS-F} spacecraft and operations (Section 2), the flare spectra obtained in its operating lifetime including a list of the lines observed (Section 3), and a description of the forthcoming {\em ChemiX} instrument due to fly on the {\em Interhelioprobe} missions (Section 4).

\section{CORONAS-F and the DIOGENESS instrument}

\subsection{Spacecraft and performance}

{\em CORONAS-F} was the second of three Russian spacecraft designed to observe various solar phenomena over the period 1994--2009, with thirteen instruments studying global oscillations, variations in ultraviolet, X-ray, and gamma-ray emission, and particle emission. The spacecraft was launched into a near-circular orbit around the Earth with an altitude range of 501--549~km. The orbital plane was inclined at $82^\circ .5$ to the equator, and its initial period was 94.9 minutes. Intervals of uninterrupted solar observation were possible for up to 20 days on two occasions during a year, while at other times the spacecraft night-time periods lasted no more than 35~minutes. Solar X-ray observations were also interrupted by passages through the auroral oval particle zones near the Earth's magnetic poles and the South Atlantic Anomaly. Spacecraft pointing at the Sun was achieved with a 15-arcminute precision.

The instrument package is described by \inlinecite{ora02}. The {\em RESIK} and {\em DIOGENESS} instruments were built by teams led by the Space Research Centre, Polish Academy of Sciences, in Wroc{\l}aw, Poland. {\em RESIK} has been described by \inlinecite{jsyl05}. It was a bent crystal spectrometer with four channels observing X-ray flare and non-flaring active region in the range 3.3--6.1~\AA, which included emission lines of highly ionized Si, S, Cl, Ar, and K ions. An absolute flux calibration of $\sim 20$\% precision was achieved for the instrument, allowing element abundances to be determined. These estimates have been discussed in several publications: see \inlinecite{bsyl14} and references therein. {\em RESIK} operated from the time of spacecraft launch until May 2003 when there was an instrument power supply failure.

\subsection{DIOGENESS: instrument and performance}

The {\em DIOGENESS} instrument by contrast with {\em RESIK} had a lifetime of only a few weeks, but the instrument's concept including that of the Dopplerometer was adequately verified. The construction of the instrument was outlined by \inlinecite{plo02} and \cite{sia02}. {\em DIOGENESS} had four channels covering wavelength ranges in the vicinity of intense lines of highly ionized Si, S, and Ca lines visible in solar flares. Two quartz crystals covered the \caxix\ lines (channels 1 and 4), and ADP and beryl crystals (channels 2 and 3) covered the \sxv\ and \sixiii\ lines respectively. Table~\ref{diog_channels} gives details of the wavelength ranges and crystals. The mono-crystals were flat and mounted on a common rotatable shaft, so a scanning motion back and forth of the shaft allowed X-rays with various wavelengths $\lambda$ to be diffracted according to Bragg's diffraction law,

\begin{equation}
n\lambda = 2d\,\, {\rm sin}\, \theta
\end{equation}

\noindent where $d$ is the crystal lattice spacing, $\theta$ the angle of incidence, and $n$ the diffraction order. Only first-order ($n=1$) spectra were recorded. The angular range of the crystal shaft was 140~arcmin, giving the wavelength ranges specified in Table~\ref{diog_channels}. The spectral resolution of each crystal is defined by the rocking curve, values of which (FWHM) are given in arcsec in Table~\ref{diog_channels}. The instrument was uncollimated but as flares generally have small angular extents (typically a few arcmin) the emission region can be considered to be a point source so a collimator is generally not necessary. The diffracted X-rays are registered by double proportional counters with beryllium entrance windows of 145~$\mu$m thickness and filled with argon gas with small admixture of carbon dioxide quenching gas at $\sim 0.5$ atmospheric pressure. An Fe$^{55}$ source (energy 5.9~keV) illuminates the other, non-solar proportional counter section so that the detector's electronic system's energy gain could be checked. A multi-slit collimator with soft X-ray detectors scanning in the direction of the spectrometer's dispersion was also included to give spatial context information, but unfortunately this part of the instrument failed early in the mission.

\begin{table}
\caption{{\em DIOGENESS} instrument channels}
\label{diog_channels}
\begin{tabular}{llcccr}
  \hline                   
Channel & 1 & 2 & 3 & 4 \\
  \hline
Crystal & Quartz & ADP & Beryl & Quartz \\
Diffracting plane & 10$\bar{1}$1 & 101 & 1010 & 10$\bar{1}$1 \\
$2d$ spacing (\AA)\tabnote{Prelaunch measurement.} & 6.6855 & 10.5657 & 15.9585 & 6.6875 \\
Central wavelength (\AA) & 3.1781 & 5.0374 & 6.6488 & 3.1781 \\
Principal lines in range & \caxix\ & \sxv\ & \sixiii\ & \caxix\ \\
Min. of wavelength range (\AA) & 3.1436 & 4.9807 & 6.1126 & 2.9601 \\
Max. of wavelength range (\AA) & 3.3915 & 5.3721 & 6.7335 & 3.2123 \\
Crystal reflectivity ($\mu$rad)$^1$ & 91 & 91 & 15 & 90 \\
Rocking curve (FWHM, arcsec)$^1$ & 24.1 & 68.1 & 94.1 & 25.6 \\
  \hline
\end{tabular}
\end{table}

\begin{figure}
\centerline{\hspace*{0.015\textwidth}
               \includegraphics[width=0.525\textwidth,clip=]{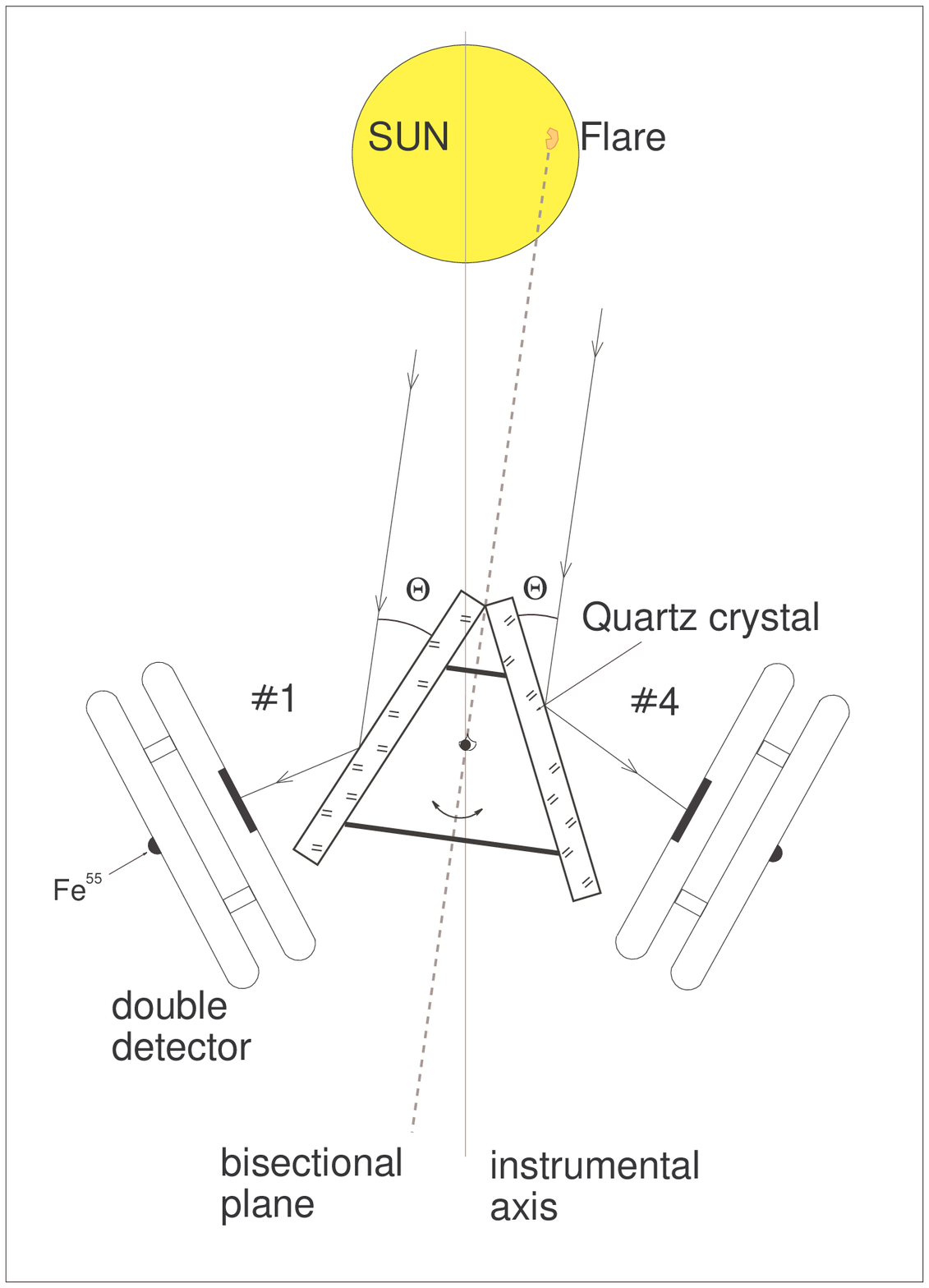}
               \hspace*{0.03\textwidth}
               \includegraphics[width=0.47\textwidth,clip=]{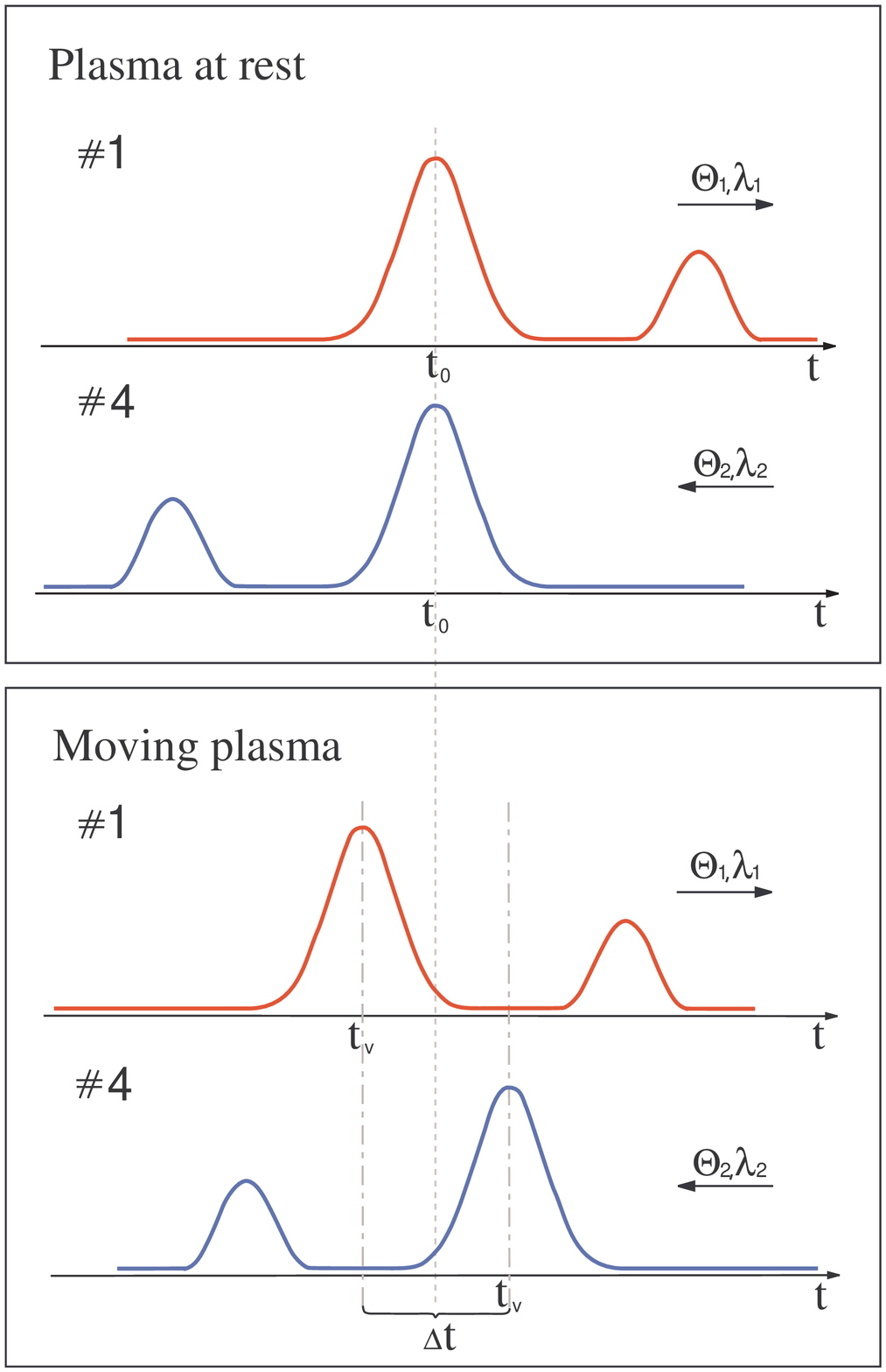}
              }
%
\caption{Left panel: Principle of {\em DIOGENESS} dopplerometer. Solar X-rays are incident on two identical quartz crystals oriented as shown. The diffracted rays around the wavelengths of the \caxix\ lines near 3.177~\AA\ are detected by double proportional counters (labelled 1 and 4 in the figure). The detector energy gain is checked with Fe$^{55}$ sources at the rear of each detector. Right panel: Resulting \caxix\ spectra, showing the more intense resonance ($w$) and less intense intercombination ($x,y$) lines (see Table~\ref{line_ids}). In the upper panel the flare plasma is assumed to be at rest with respect to the instrument, and the \caxix\ $w$ (resonance) lines are coincident (for detector 1 wavelength $\lambda$ increases to the right, for detector 4 it increases to the left). In the lower panel the flare plasma is moving with approaching velocity (as is usually the case with flare impulsive phases), so the line wavelengths are displaced to shorter wavelengths in opposite directions. Any shift due to spatial movement across the line of sight will be seen as wavelength shifts in identical directions. }\label{diag_principle}
\end{figure}

For channels 2 and 3, the \sixiii\ and \sxv\ lines were scanned over the period of a flare in the direction of alternately increasing and decreasing wavelengths repeatedly. Figure~\ref{diag_principle} shows the arrangement for channels 1 and 4 which viewed the \caxix\ lines with identical quartz crystals cut from the same parent mono-crystal at an angle $\alpha$ between the diffracting planes of the crystals, the so-called dopplerometer mode. Solar X-rays are incident on each crystal in the manner illustrated in the figure. As the two crystals on their common shaft are rotated, solar X-ray lines are successively recorded by each channel from the point-like flare on the Sun, as shown in the right-hand panel of Figure~\ref{diag_principle}. The angle $\alpha$ was selected such that the \caxix\ resonance line $w$ emitted by flare emission with zero radial velocity was observed simultaneously, its theoretical value being $56^\circ .7416$. (In practice, owing to co-alignment uncertainty, the angle $\alpha$ was set to $56^\circ .8000$.) In Figure~\ref{diag_principle}, the two occasions when the vicinity of the \caxix\ $w$ line was scanned simultaneously in the sense of increasing wavelengths (red scan in the figure) and decreasing wavelengths (blue scan) are shown. The horizontal axis indicates the stepper drive's address rather than wavelength. For zero radial velocity of the flare source, the \caxix\ $w$ line would be coincident for each scan but because of the presence of an approach velocity (such wavelength shifts have been commonly attributed to ``chromospheric evaporation'' or upward convection of the emitting plasma) the lines are separated in each scan. The linearity of the scanning mechanism was checked during the spacecraft mission using strain-gauge systems.

\begin{figure}
\centerline{\includegraphics[width=0.9\textwidth,clip=]{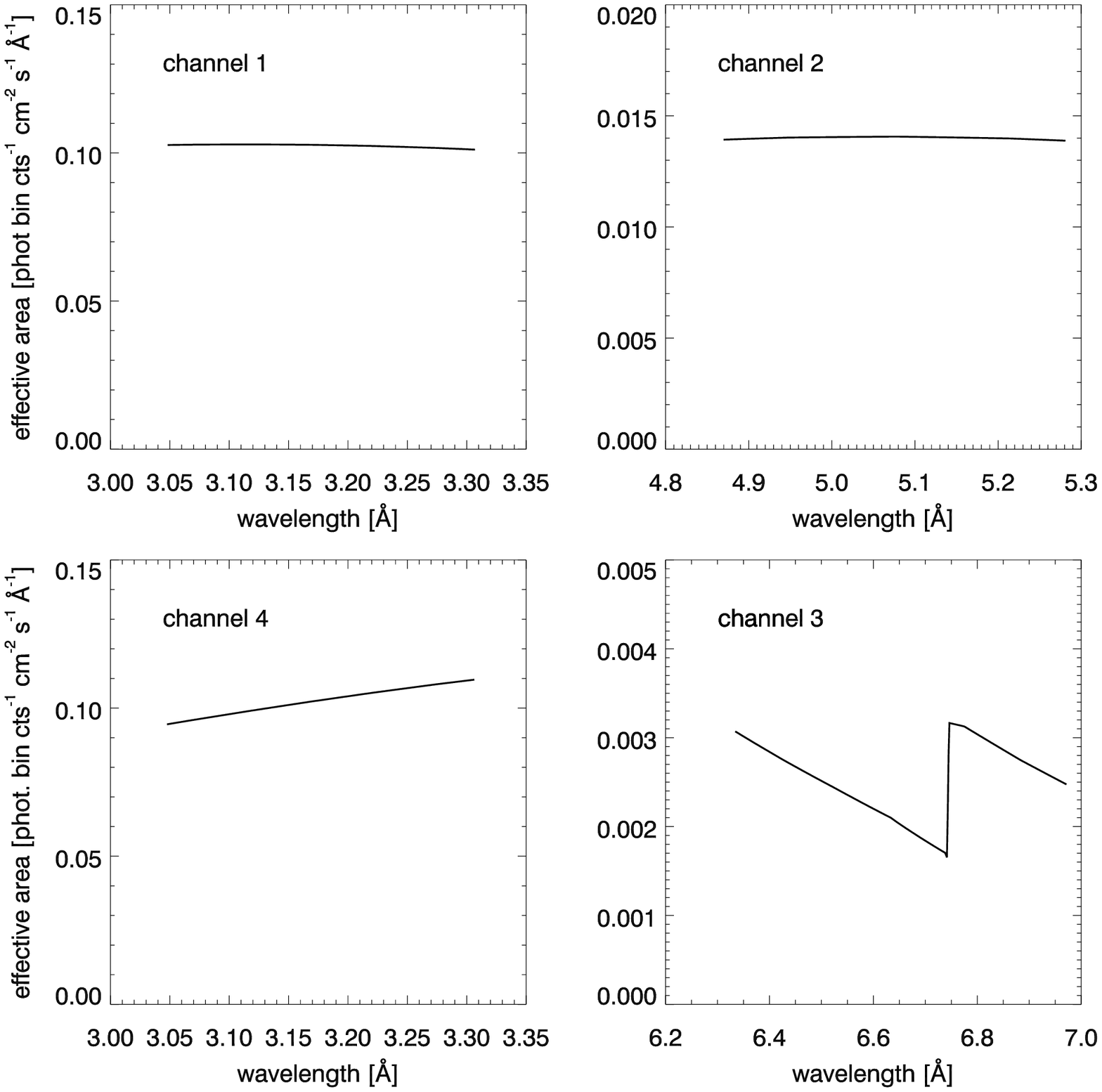}}
\caption{Effective areas of the four channels of {\em DIOGENESS}. The units are photons bin count$^{-1}$ cm$^{-2}$ sec$^{-1}$ \AA$^{-1}$. The channel numbers are indicated in the legend.}\label{eff_areas}
\end{figure}

The effective areas of the four {\em DIOGENESS} channels (see Figure~\ref{eff_areas}) were measured before spacecraft launch, and are the product of detector efficiency, filter transmission, crystal integrated reflectivity, and projected area of the detector window as seen from the Sun.

\section{ DIOGENESS Observations}

\subsection{Flare Spectra}

Eight intense flares were observed during the lifetime of {\em DIOGENESS}; see Table~\ref{flares} for details of these flares including  times, standard IAU flare notation, {\em GOES} and H$\alpha$ importance, and heliographic location. One hundred and forty spectra were obtained in the four channels for these events.

\begin{table}
\caption{Flares observed by {\em DIOGENESS}}
\label{flares}
\begin{tabular}{llclcc}
\hline                   
Date & UT of & Flare notation & {\em GOES} & H$\alpha$ & Location\tabnote{No H$\alpha$ records for Sept. 3 flares; locations from Yohkoh HXT images.}\\
(2001)& peak && importance & importance \\
\hline
Aug. 25 & 16:45 & SOL2001-08-25T16:45 & X5.3 & 3B & S17E34 \\
Aug. 30 & 17:57 & SOL2001-08-30T17:57 & M1.5 & 2N & S21W28 \\
Sept. 2 & 06:02 & SOL2001-09-02T06:02 & M1.3 & 1F & S17W66 \\
Sept. 2 & 13:48 & SOL2001-09-02T13:48 & M3.0 & 2N & S21W65 \\
Sept. 3 & 01:58 & SOL2001-09-03T01:58 & C9.0 & -  & S17E90 \\
Sept. 3 & 17:16 & SOL2001-09-03T17:16 & M1.1 & -  & N10W06 \\
Sept. 3 & 18:41 & SOL2001-09-03T18:41 & M2.5 & -  & S26E90 \\
Sept. 16 & 03:53 & SOL2001-09-16T03:53 & M5.6 & 2N & S29W54 \\
\hline
\end{tabular}
\end{table}

\begin{figure}
\centerline{\includegraphics[width=0.8\textwidth,clip=,angle=90]{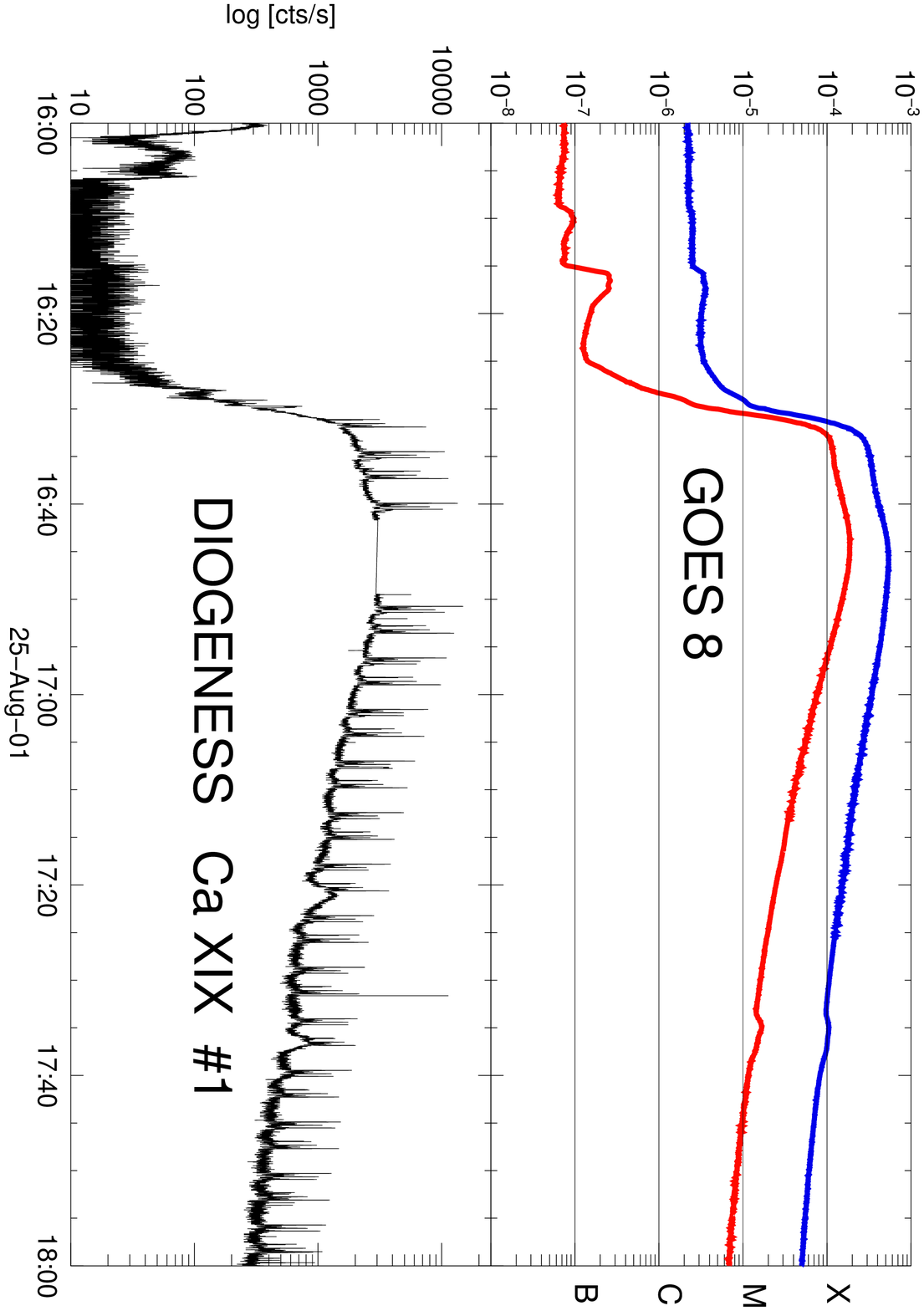} }
\caption{Top: Light curves of the two channels of {\em GOES} (red - 0.5--\AA; blue - 1--8\AA), irradiance scale in W~m$^{-2}$ on left axis ({\em GOES} importance scale on right axis) for the 25~August 2001 flare. Lower: {\em DIOGENESS} \caxix\ count rate (logarithmic scale), showing the scans through the \caxix\ lines as the crystals rocked back and forth (time of scan 140~s). A passage of the {\em CORONAS-F} spacecraft through the auroral oval radiation belts before the flare occurred around 16:00--16:10~UT.  }\label{GOES_diag_lightcurves_color}
\end{figure}

\begin{figure}
\centerline{\hspace*{0.015\textwidth}
               \includegraphics[width=0.49\textwidth,clip=]{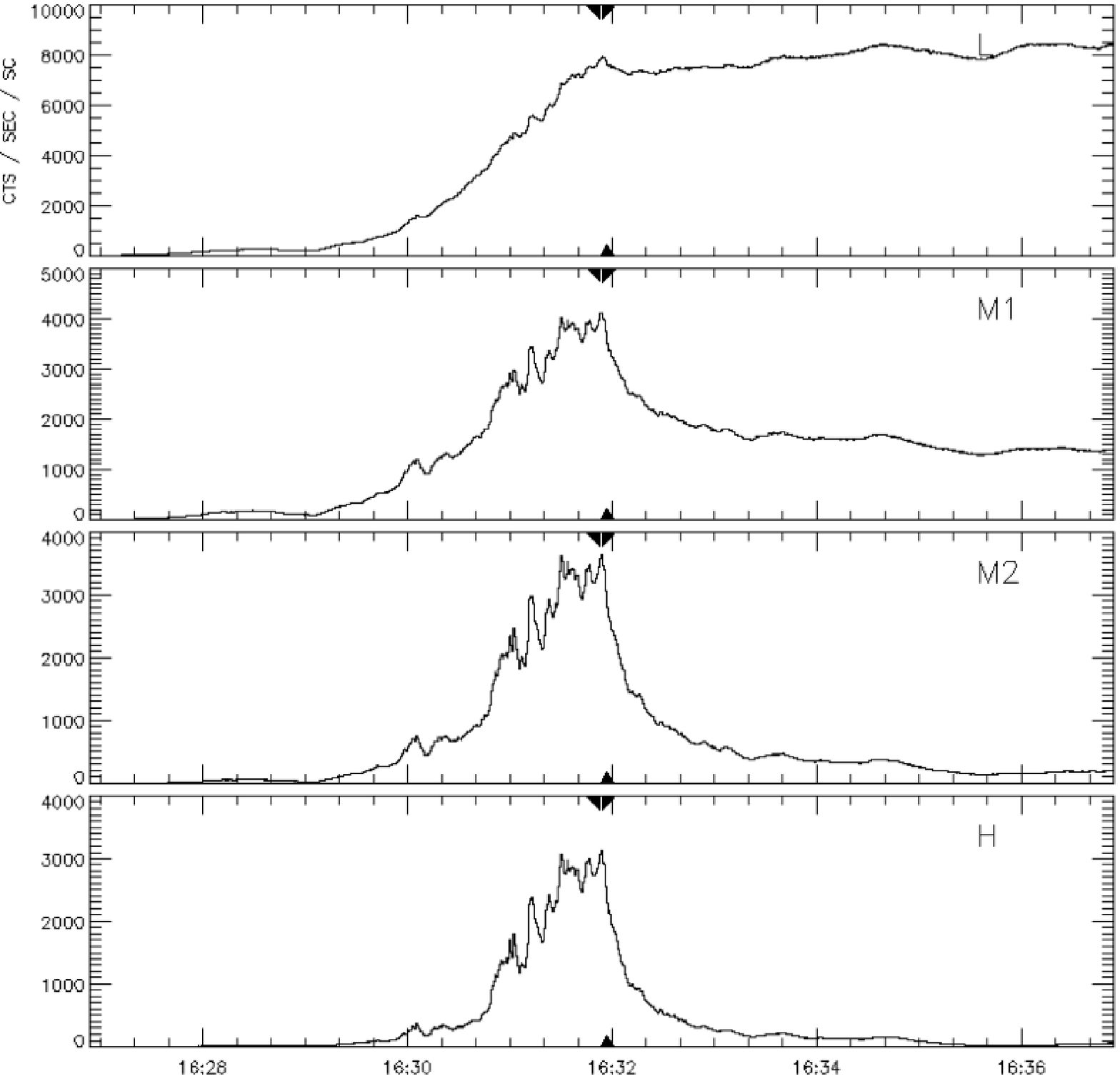}
               \hspace*{0.03\textwidth}
               \includegraphics[width=0.49\textwidth,clip=]{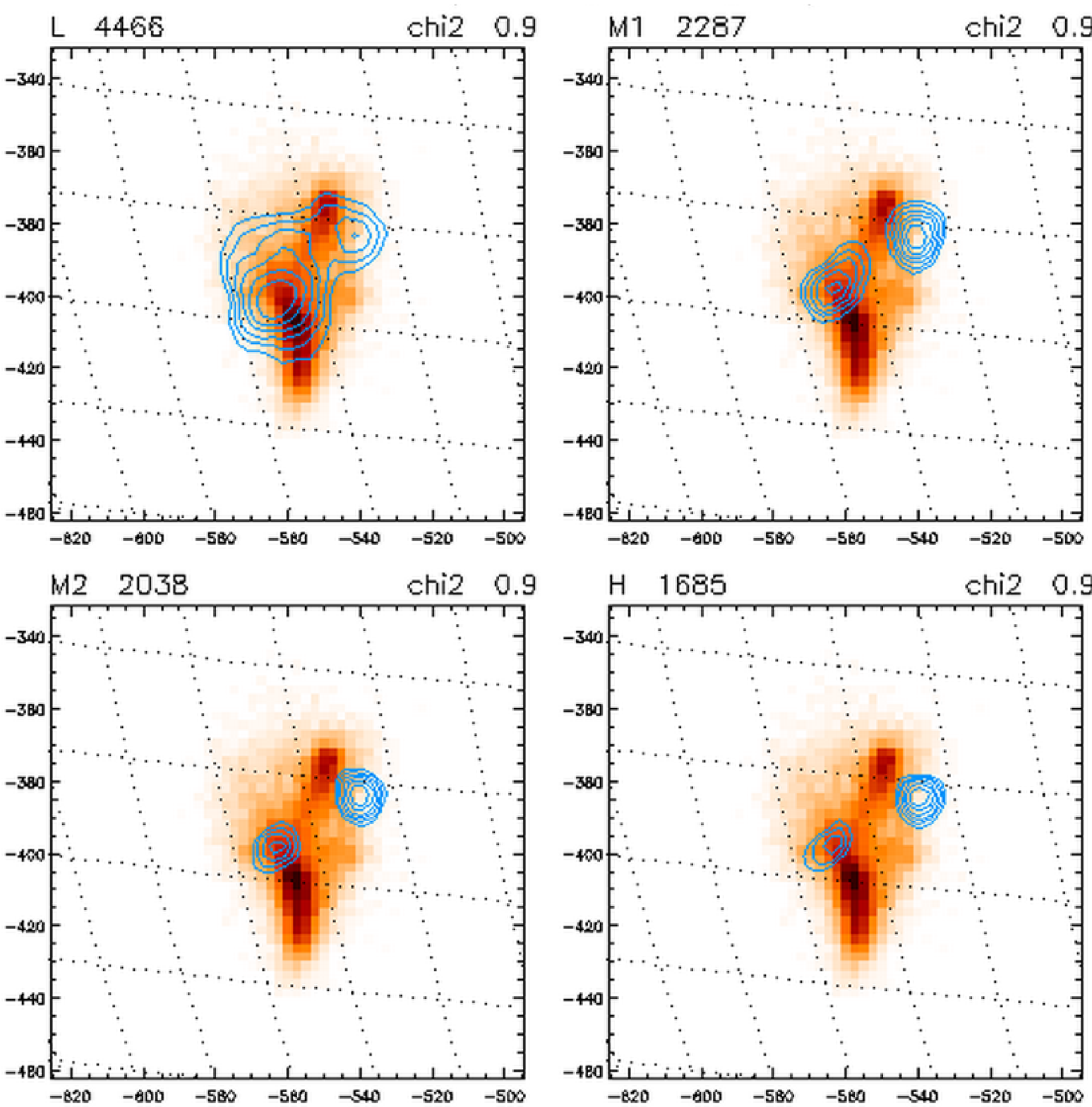}
              }
\vspace{-0.35\textwidth}   
     \centerline{\Large \bf     
      \hspace{0.0 \textwidth}  \color{white}{(a)}
      \hspace{0.415\textwidth}  \color{white}{(b)}
         \hfill}
     \vspace{0.31\textwidth}    
\caption{Left: Light curves over the impulsive stage of the 25 August 2001 flare from the {\em Yohkoh} HXT instrument in its four channels: L - 14-23~keV, M1 - 23-33~keV, M2 - 33-53~keV, H - 53-93~keV. Right: {\em Yohkoh} SXT images (red intensity scale) and HXT images (blue contours) in the four channels of HXT indicated at the top of each image. }\label{hxt_Aug25flare}
\end{figure}

The X5 flare on 25~August 2001 was observed to great advantage by {\em DIOGENESS}, from the earliest pre-flare state including a small precursor flare (16:17~UT) and the impulsive stage at around 16:30--16:32~UT, also seen with the {\em Yohkoh} Hard X-ray Telescope (HXT), until late in the decay stage, a total of nearly two hours. Figure~\ref{GOES_diag_lightcurves_color} shows the light curves from the two {\em GOES} channels and the total emission in {\em DIOGENESS} channel~1 (\caxix\ channel). The {\em DIOGENESS} light curve shows the \caxix\ line emission as the Bragg diffraction condition (Eq.~1) was successively satisfied for the \caxix\ emission lines a total of $\sim 30$ scans, i.e. approximately 140~s per scan. At the impulsive stage, the HXT saw emission up to its highest-energy channel (53--93 keV): see the light curves at four energy bands in Figure~\ref{hxt_Aug25flare} (left-hand panel). The HXT images show the double-footpoint emission characteristic of large flares with soft X-ray emission (from the {\em Yohkoh} Soft X-ray Telescope) in between (see right-hand panel of Figure~\ref{hxt_Aug25flare}).

\begin{figure}
\centerline{\hspace*{0.015\textwidth}
               \includegraphics[width=0.49\textwidth,clip=]{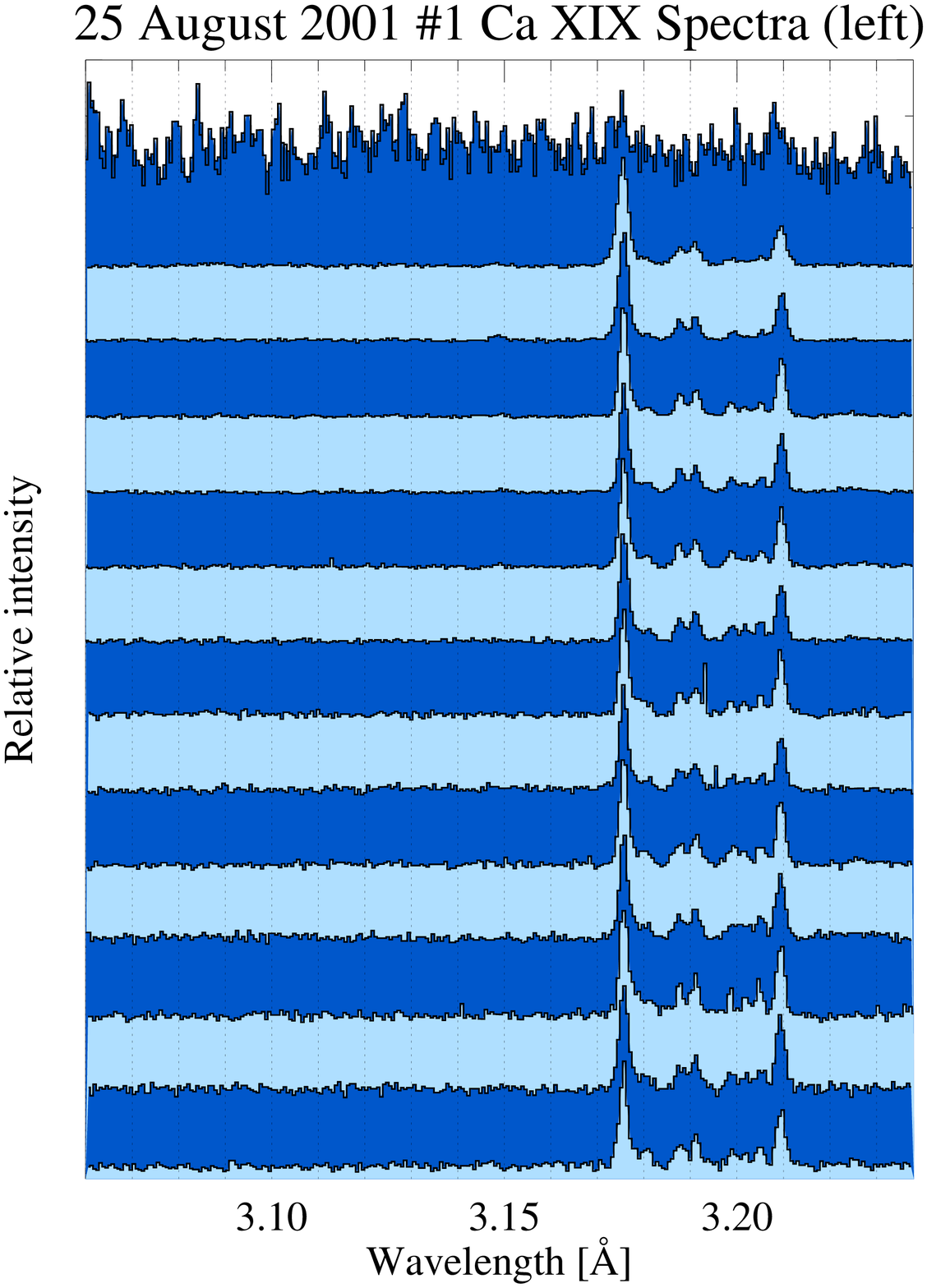}
               \includegraphics[width=0.5\textwidth,clip=]{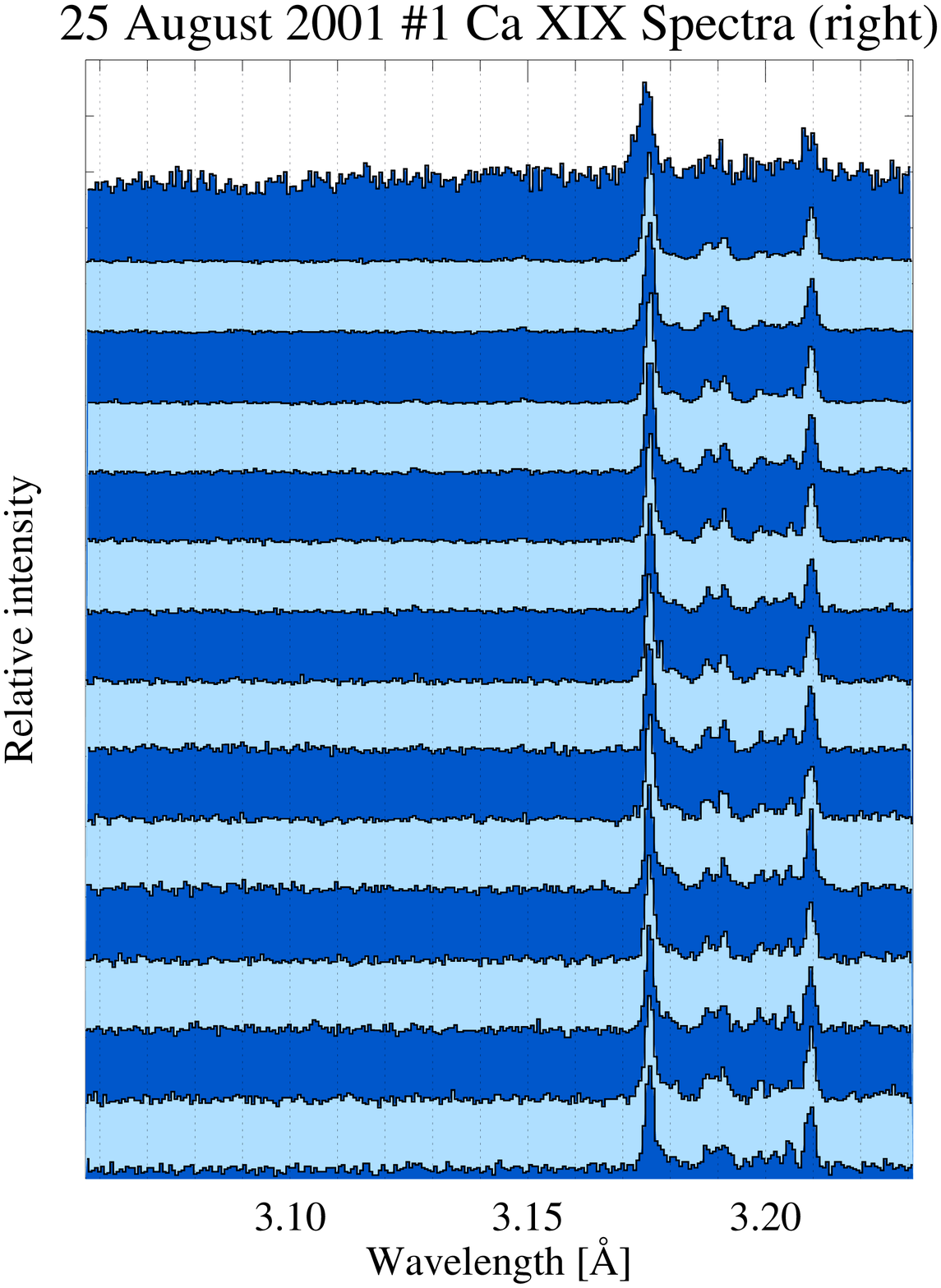}
 }
\caption{Left: Forward (increasing wavelength) scans for {\em DIOGENESS} channel 4 (\caxix\ lines) during the 25~August 2001 flare stacked with increasing times from top to bottom (time range 16:29--18:00~UT). Right: Corresponding backward (decreasing wavelength) scans. }  \label{caxix_scans}
\end{figure}

In Figure~\ref{caxix_scans}, channel 4 spectra over the \caxix\ lines are indicated during the 25~August 2001 flare with forward and backward scans (wavelength increasing and decreasing with time respectively).

\subsection{Plasma Dynamics in the 25 August 2001 Flare}

\begin{figure}
\centerline{\hspace*{0.015\textwidth}
\includegraphics[width=0.55\textwidth,clip=,angle=0]{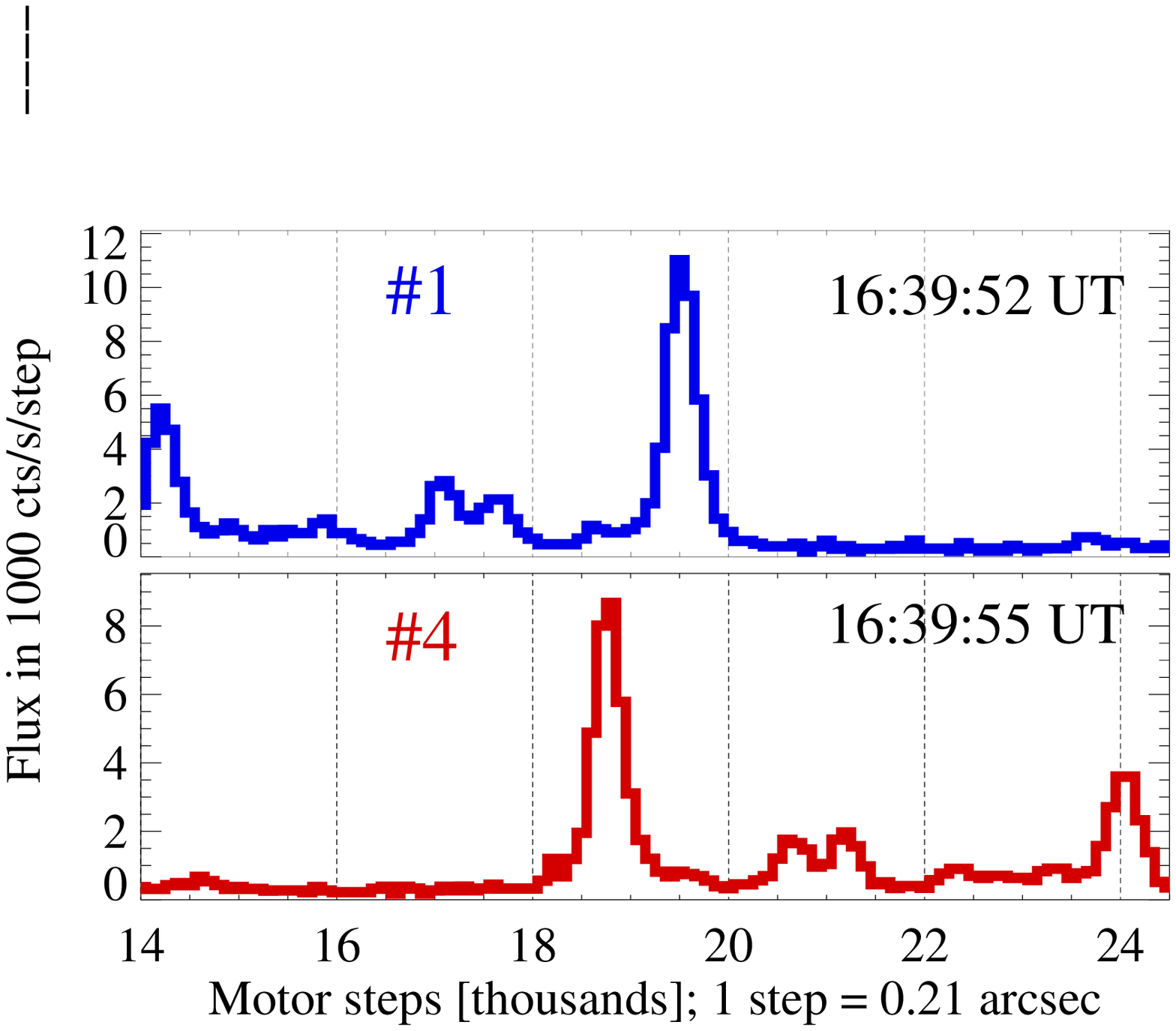}
\includegraphics[width=0.45\textwidth,clip=,angle=0]{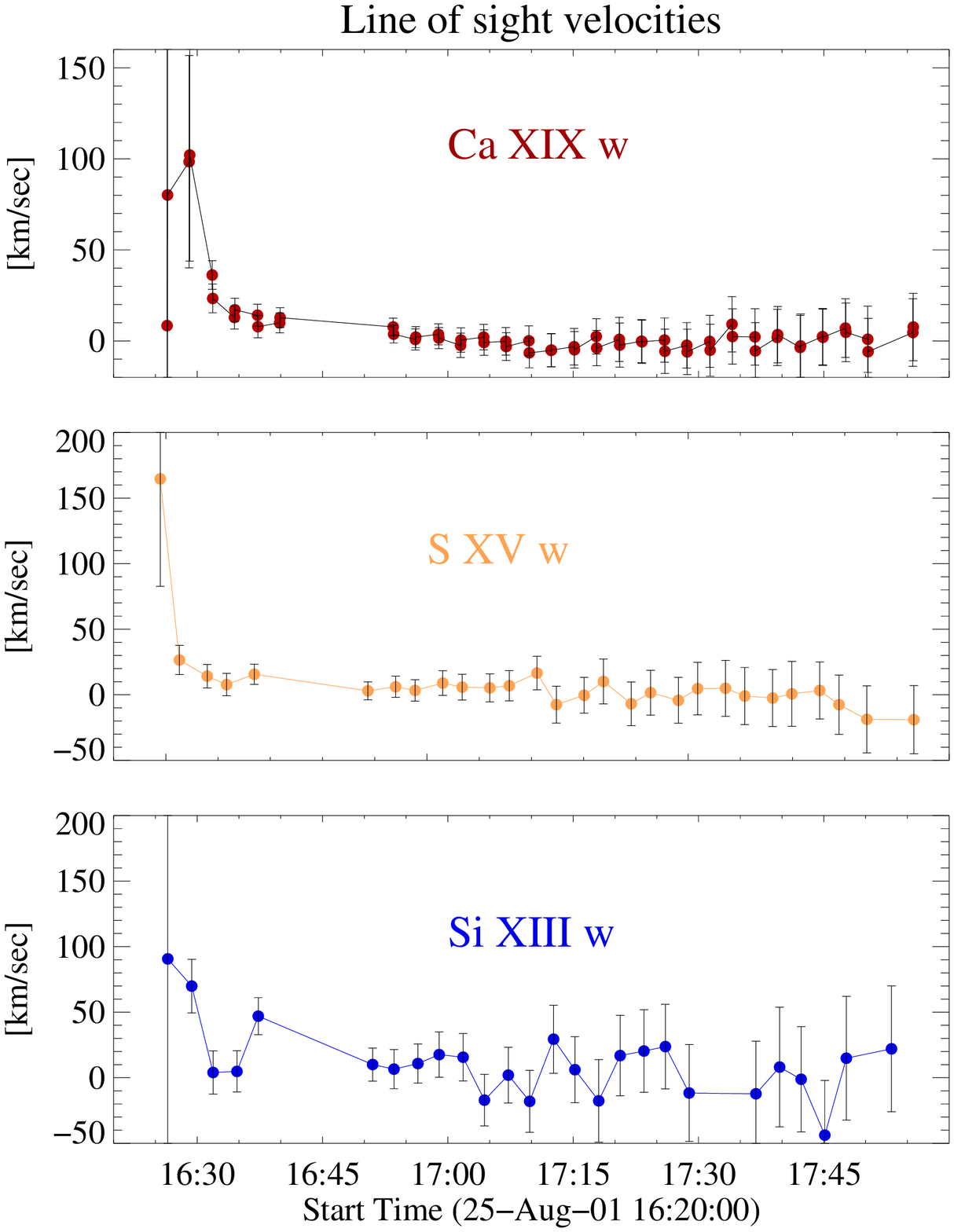}}
\caption{Left: \caxix\ spectra from channels 1 and 4 (reversed), plotted as a function of stepper motor addresses, for two times (indicated) in the 25~August 2001 flare. Right: Line-of-sight velocities of X-ray flare plasma as deduced from Doppler shifts of \caxix, \sxv, and \sixiii\ lines during the flare of 25~August 2001. See \cite{plo02} for details.}\label{diog_vels}
\end{figure}

Velocities of the emitting plasma during the 25 August 2001 flare were found from {\em DIOGENESS} \sixiii, \sxv, and \caxix\ spectra, with the \caxix\ spectra from the dopplerometer part of the instrument. The left panel of Figure~\ref{diog_vels} shows the \caxix\ spectra in channels 1 and 4 plotted in stepper motor addresses. The main lines apparent are the resonance ($w$), intercombination ($x$, $y$), and forbidden ($z$) lines of \caxix\ in order of wavelength (see Table~\ref{line_ids}). The channel~1 spectrum (in blue in Figure~\ref{diog_vels}) has wavelength increasing to the left, the channel~4 spectrum (in red) with channel~1 to the right. The scan times of the spectra are almost coincident. The deduced velocities for all channels are given as a function of time in Figure~\ref{diog_vels} (right panel). They indicate that there were high approach velocities at the time (approximately 16:31~UT) of the hard X-ray impulses recorded by the {\em Yohkoh} HXT M1, M2, and H channels (see light curves and images in Figure~\ref{hxt_Aug25flare}). The highest velocity, 150~km~s$^{-1}$, is recorded by the \sxv\ (temperature $\sim 10$~MK) line emission. The velocity of the higher-temperature ($\sim 20$~MK) \caxix\ line emission indicates a velocity of about 100~km~s$^{-1}$ as does the lower-temperature ($\sim 7$~MK) \sixiii\ line emission.

These velocities are rather smaller than those previously observed for large disk flares. Thus, \cite{dos80} in discussing several X-class flares seen with X-ray spectrometers on the {\em P78-1} spacecraft found approach velocities of about 400~km~s$^{-1}$ for disk flares, as did \cite{tan82} for a very large disk flare seen with an X-ray spectrometer on the {\em Hinotori} spacecraft. It is possible that plasma motions during the 25~August 2001 flare observed by {\em DIOGENESS} were directed at a substantial angle to the line of sight (although the flare was on the disk, it was $34^\circ$ east of the central meridian), so accounting for the smaller observed approach velocities in its rise phase.

%

\subsection{Spectral Line Identifications}

\begin{figure}
\centerline{\includegraphics[width=0.7\textwidth,clip=,angle=90]{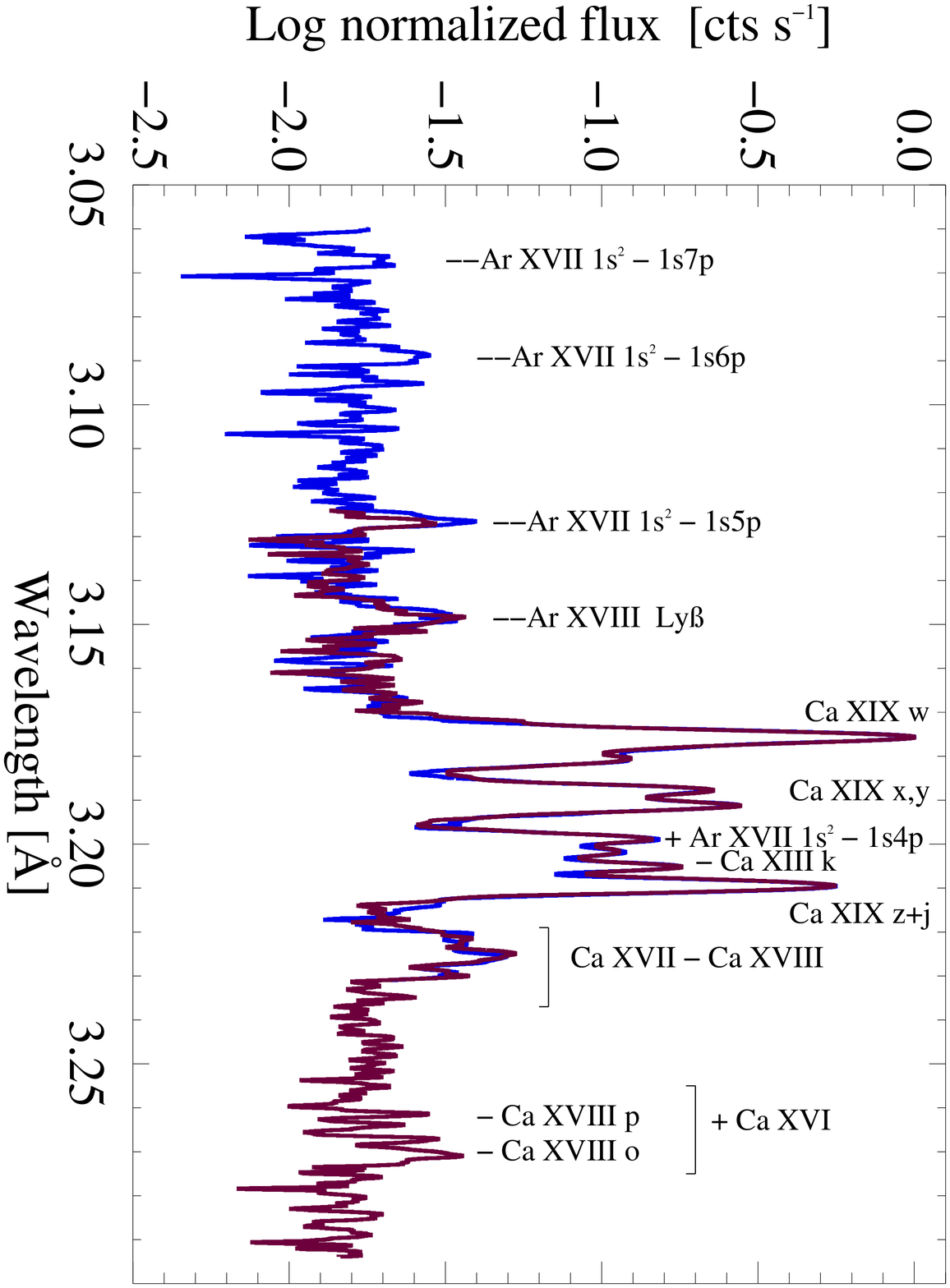}}
\caption{Averaged {\em DIOGENESS} spectrum from channels~1 and 4 during the 25~August 2001 flare showing \arxvii\ and \caxix\ lines and dielectronic satellites of \caxviii, \caxvii, and \caxvi. }\label{diog_sp}
\end{figure}

Figure~\ref{diog_sp} shows an averaged spectrum from {\em DIOGENESS} channels~1 and 4 during the 25~August 2001 flare, peak time in soft X-rays at 16:45~UT (Table~\ref{flares}), with flux on a logarithmic scale to show the weak line emission either side of the \caxix\ line emission to better advantage. Table~\ref{line_ids} gives the wavelengths of identifiable line features for the $3.05-3.30$~\AA\ range. Channels~1 and 4 spectra cover the region of the \caxix\ resonance ($w$) line and includes other \caxix\ lines and dielectronic satellites of the Li-like (\caxviii) and lower stages of ionization. These lines are familiar from previously flown X-ray spectrometers on the {\em P78-1}, {\em SMM}, {\em Hinotori}, and {\em Yohkoh} spacecraft. The ratios of the principal satellites ($k$ and the $d13$, $d15$ lines) to the \caxix\ $w$ line have been used extensively to find flare and non-flaring active region temperatures and emission measures, based on the theory of \inlinecite{gab72} (whose line notation we use here) and \inlinecite{bel82}. \caxviii\ satellite $j$ is indistinguishable from \caxix\ line $z$. The ratio of the inner-shell excitation satellite $q$ to line $w$ depends on the ion fraction $N({\rm Ca}^{+17}) / N({\rm Ca}^{+18})$ and so is in principle a means of finding departures of ionization equilibrium in the emitting plasma. However, satellite $q$ is blended with the \arxvii\ line $w4$ (this notation being used for the transition $1s^2 - 1s4p$).

In the {\em DIOGENESS} spectrum shown in Figure~\ref{diog_sp}, the Rydberg series of \arxvii\ lines up to the $w7$ line can be discerned. The \caxix\ and \arxvii\ spectra up to the $w10$ line have been observed during strong flares with crystal spectrometers on the {\em P78-1} spacecraft \citep{dos85,see85}, with lines $w9$ and $w10$ blended with \caxix\ dielectronic satellites on the long-wavelength side of the \caxx\ Ly-$\alpha$ doublet. Spectra from the Alcator~C-Mod tokamak plasmas, recently reported by \cite{rice14}, also show \arxvii\ high-$n$ lines up to the ionization limit at 3.009~\AA. All these lines, which are shortward of 3.06~\AA, are outside the {\em DIOGENESS} range. We note that the estimated temperatures ($\sim 30$~MK) and electron densities ($\sim 10^{15}$~cm$^{-3}$) in the Alcator plasmas are both higher than those typical of large solar flare plasmas ($\sim 25$~MK and $\sim 10^{11}-10^{12}$~cm$^{-3}$ respectively), but this has little effect on the appearance of the spectral lines, the principal excitation mechanism for which is electron collisional excitation from the ground state of ions in both cases. 

In Table~\ref{line_ids}, we follow \inlinecite{rice14} and \inlinecite{see89} in identifying lines in the {\em DIOGENESS} spectrum on the long-wavelength side of the \caxix\ forbidden line $z$ as \caxvii\ satellites including the strong inner-shell-excited line $\beta$, as well as \caxviii\ satellites $u$ and $v$ which are both dielectronically and collisionally excited \citep{gab72}. In addition, a feature possibly made up of several lines between 3.261~\AA\ and 3.272~\AA, not apparently identified before, is likely to be due to a group of \caxvi\ lines by analogy with equivalent lines of \fexxii\ seen in solar flare spectra with the {\em SMM} Bent Crystal Spectrometer \citep{phi83}. The \caxviii\ two-electron-jump satellites $o$ and $p$ (transitions $1s^2\,2p\,^2P_{3/2} - 1s\,2s^2\,^2S_{1/2}$, $1s^2\,2p\,\,^2P_{1/2} - 1s\,2s^2\,\,^2S_{1/2}$) may also contribute to this line feature: their wavelengths are estimated by \inlinecite{bel82} to be 3.2688~\AA\ and 3.2636~\AA\ respectively.

In {\em DIOGENESS} channel~2 spectra, the predominant lines are the resonance, intercombination, and forbidden lines of \sxv\ and the Ly-$\beta$ ($1s-3p$) line of \sixiv, prominent also in RESIK spectra. The $x$ and $y$ lines are partially blended as are \sxiv\ satellites $d13$, $d15$ with \sxv\ line $w$ and \sxiv\ satellites $j$, $k$ with \sxv\ line $z$. There is a possible long-wavelength ``shoulder'' of emission to the \sixiv\ line which may be due to the \sixiii\ $w6$ line. In channel~3 spectra, the \sixiii\ $w$--$z$ lines are the most prominent with \sixii\ satellites $d13 + d15$, $q$, and $k$ identifiable. Possible very weak line features to the short-wavelength side of the \sixiii\ line $w$ occur. The \mgxii\ Rydberg sequence $1s-np$ occur in this region, and were seen with the crystal spectrometers on {\em OSO-8} \citep{par78}. They may account for the 6.410~\AA\ ($n=8$), 6.490~\AA\ ($n=6$), and 6.580~\AA\ ($n=5$) lines in the {\em DIOGENESS} spectra although the $n=7$ \mgxii\ line at 6.448~\AA\ is not apparent. The $n=4$ \mgxii\ line blends with the \sixiii\ $z$ line. Note that the anomaly in the effective area due to the beryl crystal reflectivity at about 6.75~\AA\ (see Figure~\ref{eff_areas}) distorts the intensity of the \sixiii\ $z$ line.

In time, the line identifications given in Table~\ref{line_ids} will be refined with better atomic data. In particular, we plan to run atomic codes to find the wavelengths of the satellites due to \caxvi\ as was done for the case of the equivalent Fe ions \citep{phi83}.

\begin{table}
\caption{Spectral line identifications in {\em DIOGENESS} spectra}
\label{line_ids}
\begin{tabular}{cccccc}
  \hline                   
\multicolumn{2}{c}{{\em Channels 1 and 4}} \\
Measured & \multicolumn{4}{c}{Line identification} \\
line $\lambda$ (\AA) & Ion & $\lambda$ (\AA) & Transition & Name\tabnote{Gabriel(1972) for He-like ion $1s^2-1s2l$ transitions and Li-like ion satellites.}& Source\tabnote{D=Doschek et al. (1985); K=Kelly (1987); R=Rice et al. (2014); Ph=Phillips et al. (1983); S=Seely \& Feldman (1985).} \\
3.068 & \arxvii\ & 3.068 & $1s^2\,^1S_0 - 1s7p\,^1P_1$ & $w7$ & R,S \\
3.089 & \\
3.095 &  \arxvii\ & 3.095 & $1s^2\,^1S_0 - 1s6p\,^1P_1$ & $w6$ & K,S \\
3.127 &  \arxvii\ & 3.128 & $1s^2\,^1S_0 - 1s5p\,^1P_1$ & $w5$ & K,S \\
3.148 & \\
3.177 & \caxix\ & 3.177 & $1s^2\,^1S_0 - 1s2p\,^1P_1$ & $w$ & K \\
3.181 & \caxviii\ & 3.1809 & $1s^2 3p\, ^2P_{3/2} - 1s2p (^1P)3p\, ^2D_{5/2}$ & $d13$ & K \\
3.188 & \caxix\ & 3.1889 & $1s^2\,^1S_0 - 1s2p\,^3P_2$ & $x$ & K \\
3.191 & \caxix\ & 3.1925 & $1s^2\,^1S_0 - 1s2p\,^3P_1$ & $y$ & K \\
3.199 & \caxviii\ & 3.2003 & $1s^2\, 2s\, ^2S_{1/2} - 1s(2s2p ^3P)\, ^2P_{3/2}$ & $q$ & K \\
&\arxvii\ & 3.200 & $1s^2\,^1S_0 - 1s4p\,^1P_1$ & $w4$ & K \\
3.203 & \caxviii\ & 3.2031 & $1s^2\, 2s\, ^2S_{1/2} - 1s(2s2p ^3P)\, ^2P_{1/2}$ & $r$ & K \\
3.206 & \caxviii\ & 3.2058 & $1s^2\, 2p\, ^2P_{1/2} - 1s2p^2\, ^2D_{3/2}$ & $k$ & K \\
3.210 & \caxix\ & 3.2111 & $1s^2\,^1S_0 - 1s2s\,^3S_1$ & $z$ & K \\
&\caxviii\ & 3.2097 & $1s^2 2p\, ^2P_{3/2} - 1s2p^2\, ^2D_{5/2}$ & $j$ & K \\
3.222 & \caxvii\ & 3.2217 & $1s^2\, 2s^2\, ^1S_0 - 1s\,2s^2\,2p\,^1P_1$ & $\beta$ & R \\
3.225 & \caxvii\ & ? & $1s^2 2s^2 - 1s 2s^2 2p$ & & Ph\\
3.225 & \caxviii\ & 3.2266 & $1s^2\, 2s\, ^2S_{1/2} - 1s2s2p\, ^4P_{3/2}$ & $u$ & R,D \\
3.230 & \caxviii\ & 3.2277 & $1s^2\, 2s\, ^2S_{1/2} - 1s2s2p\, ^4P_{1/2}$ & $v$ & R,D \\
3.230 & \caxvii\ & ? & $1s^2 2s^2 - 1s 2s^2 2p$ & & Ph\\
3.272 & \caxvi\ & ? & $1s^2 2s^2 2p - 1s 2s 2p^2$ & & Ph\\
\hline
\multicolumn{2}{c}{{\em Channel 2}} \\

5.037 & \sxv\ & 5.0385 & $1s^2\,^1S_0 - 1s2p\,^1P_1$ & $w$ & K \\
5.047 & \sxiv\ & 5.0468 & $1s^2 3p\, ^2P_{3/2} - 1s2p (^1P)3p\, ^2D_{5/2}$ & $d13$ & K \\
5.062 & \sxv\ & 5.06? & $1s^2\,^1S_0 - 1s2p\,^3P_2$ & $x$ & K \\
5.064 & \sxv\ & 5.0662 & $1s^2\,^1S_0 - 1s2p\,^3P_1$ & $y$ & K \\
5.083 & \sxiv\ & 5.0861 & $1s^2 2s\, ^2S_{1/2} - 1s(2s2p ^3P)\, ^2P_{3/2}$ & $q$ & K \\
5.099 & \sxiv\ & 5.0964 & $1s^2 2p\, ^2P_{1/2} - 1s2p^2\, ^2D_{3/2}$ & $k$ & K \\
& \sxv\ & 5.101 & $1s^2\,^1S_0 - 1s2s\,^3S_1$ & $z$ & K \\
& \sxiv\ & 5.1010 & $1s^2 2p\, ^2P_{3/2} - 1s2p^2\, ^2D_{5/2}$ & $j$ & K \\
& \sixiv\ & 5.217 & $1s\, ^2S_{1/2} - 2p\, ^2P_{1/2,3/2}$ & Ly-$\beta$ \\

\hline
\multicolumn{2}{c}{{\em Channel 3}} \\

6.410 & \mgxii\ ? & 6.497 & $1s\,^2S_{1/2} - 8p\,^2P_{1/2,3/2}$ & Ly-$\eta$ & K \\
6.490 & \mgxii\ ? & 6.497 & $1s\,^2S_{1/2} - 6p\,^2P_{1/2,3/2}$ & Ly-$\epsilon$ & K \\
6.513 & \\
6.545 & \\
6.574 & \mgxii\ ? & 6.580 & $1s\,^2S_{1/2} - 5p\,^2P_{1/2,3/2}$ & Ly-$\delta$ & K \\
6.641 & \sixiii\ & 6.6477 & $1s^2\,^1S_0 - 1s2p\,^1P_1$ & $w$ & K \\
6.681 & \\
6.686 & \sixiii\ & 6.6879 & $1s^2\,^1S_0 - 1s2p\,^3P_1$ & $y$ & K \\
6.713 & \sixii\ & 6.718 & $1s^2 2s\, ^2S_{1/2} - 1s(2s2p\, ^3P)\, ^2P_{3/2}$ & $q$ & K \\
6.737 & \sixiii\ & 6.740 & $1s^2\,^1S_0 - 1s2s\,^3S_1$ & $z$ & K \\

\hline
\end{tabular}
\end{table}

\section{The ChemiX Instrument for Interhelioprobe}

\begin{figure}
\centerline{\includegraphics[width=0.8\textwidth,clip=,angle=0]{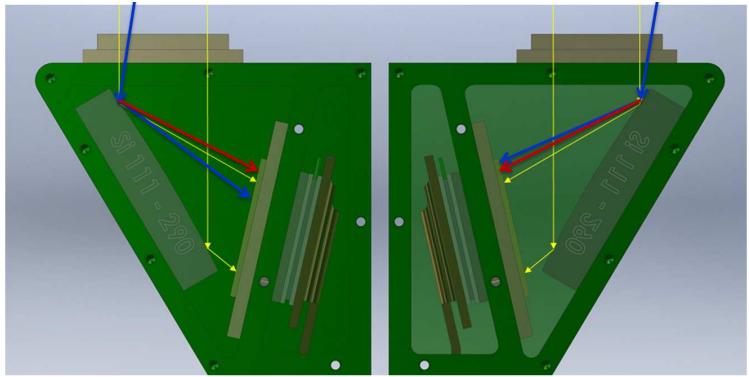}}
\caption{Scheme of ChemiX. Solar X-rays are incident from the top of the figure (blue, yellow, and red arrows), are diffracted off the two bent crystals (Si 111 in both cases) and are then incident on the CCDs. The yellow rays are for a source in the centre of the collimator field of view (FOV), the blue rays from a source moved across the FOV (tangential motion), and the red rays for a source approaching the instrument (so subject to a Doppler shift).  }\label{ChemiX_Doppler_scheme}
\end{figure}

{\em ChemiX} (CHEMical composition In X-rays) is a bent crystal spectrometer under construction that will observe the soft X-ray spectra of solar flares, continuing the objectives of previous spectrometers such as the Flat Crystal Spectrometer and Bent Crystal Spectrometer on {\em Solar Maximum Mission} \citep{act80}, SOLEX and SOLFLEX on {\em P78-1} \citep{dos79}, and {\em RESIK} and {\em DIOGENESS} on {\em CORONAS-F}. We give here a brief description while a more detailed account is in preparation (Siarkowski et al.).  {\em ChemiX} will have four ``spectrometer'' channels viewing spectral lines of \fexxv, \caxix, \sxv, \sixiii\ and \sixiv, and three pairs of crystals in a dopplerometer arrangement like that on {\em DIOGENESS} and its predecessors viewing lines of \fexxv, \caxix, and \arxvii. There will be important improvements over both the {\em RESIK} and {\em DIOGENESS} instruments in that the detectors will be X-ray-sensitive, cooled (to $<-20^\circ$ C.) charged coupled devices (CCDs) and there will be a two-dimensional collimator with $\sim 3$~arcmin (FWHM) field of view that will enable the instruments to view single active regions and so avoid possible spatial and spectral confusion for the spectrometer channels. A projected 1-s time resolution will allow the highly dynamic impulsive phases of flares to be observed, in particular changes to spectral line profiles and shifts indicative of plasma turbulence and motions. Table~\ref{chemix_channels} gives details of the Bragg-diffracting crystals to be used and the wavelength ranges in each of the seven channels. There will be full spectral coverage in the range $1.5-9$\AA\ with a spectral resolution that should be four to ten times better than that of {\em RESIK} because of the use of CCDs over gas proportional counters. There will also be much improved signal-to-noise ratio, and because of the use of crystals with low atomic number, there will be very little instrumental background due to crystal fluorescence, enabling the solar continuum to be measurable. Figure~\ref{ChemiX_Doppler_scheme} gives the scheme for {\em ChemiX} which is under construction at present.

\begin{table}
\caption{ChemiX Spectral Channels}
\label{chemix_channels}
\begin{tabular}{lccrcc}
  \hline                   
Channel & Crystal & Diffracting & $2d$ & Wavelength & Av. spectral \\
No. & & plane & (\AA) & Range (\AA) & resolution  \\
&&&&&(m\AA\ per pixel)\\
 \hline
\multicolumn{4}{c}{Spectrometer channels}\\
\hline
1 & Si & 111 & 6.271 & 1.500--2.713 & 1.46 \\
2 & Quartz & $10\bar{1}0$ & 8.514 & 2.700--4.304 & 1.95 \\
3 & KDP & 011 & 10.185 & 4.290--5.228 & 1.43 \\
4 & KAP & 001 & 26.64 & 5.200--8.800 & 4.73 \\
\hline
\multicolumn{4}{c}{Dopplerometer channels}\\
\hline
1 & LiF & 022 & 2.848 & 1.830--1.925 & 0.10 \\
2 & Si & 111 & 6.271 & 3.150--3.245 & 0.10 \\
3 & Si & 111 & 6.271 & 3.90--4.080 & 0.19 \\

  \hline
\end{tabular}
\end{table}

\begin{figure}
\centerline{\includegraphics[width=0.9\textwidth,clip=,angle=0]{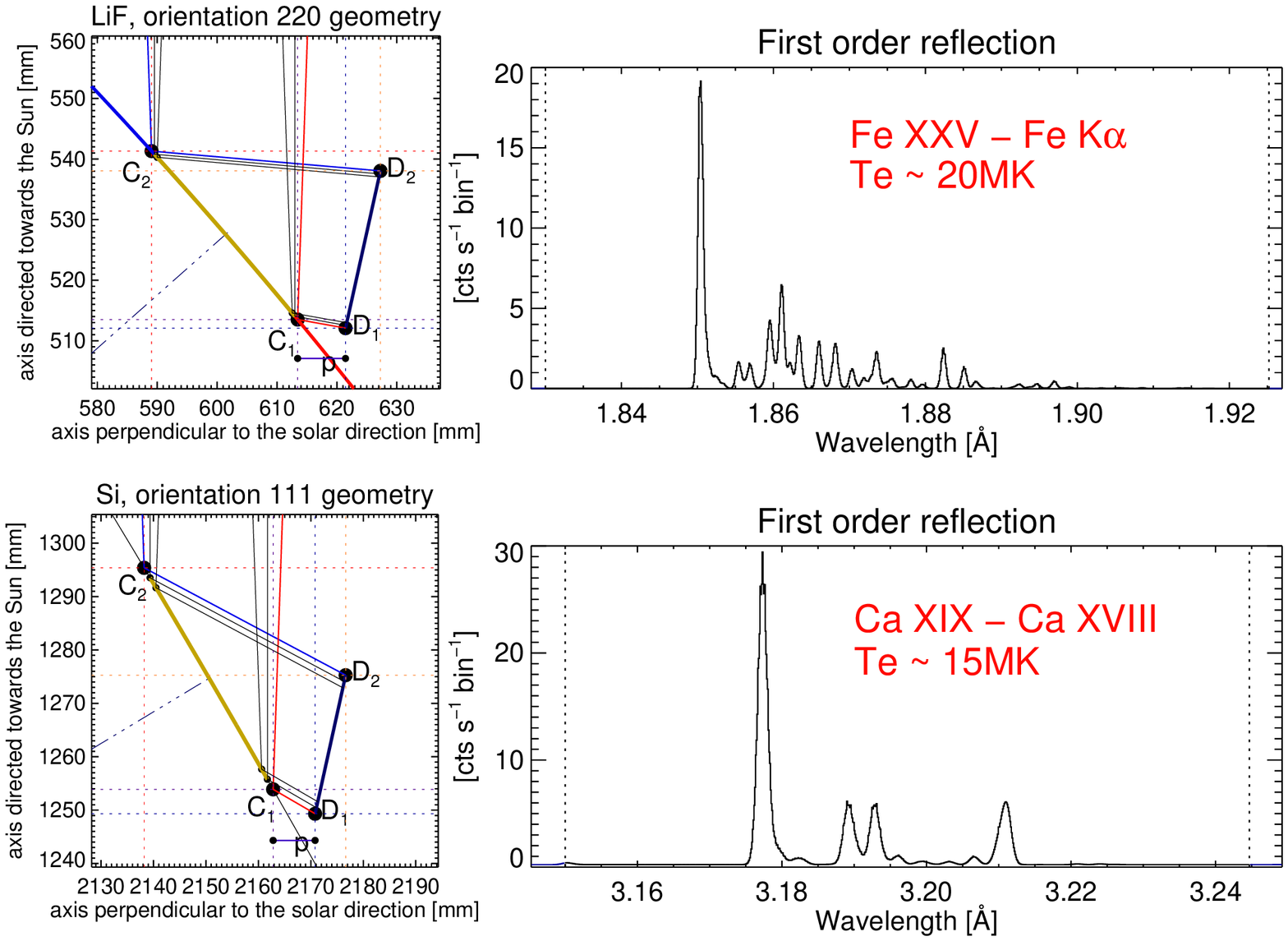}}
\caption{Scheme of crystal diffraction for dopplerometer channels 1 (\fexxv, top) and 2 (\caxix, lower) and resulting spectra (simulated for a {\em GOES} M5 flare using the {\sc chianti} database and software package). Left: Solar X-rays are incident from the top of the figure on the crystal (light green), are diffracted by the crystals ($C_1 - C_2$) and then are incident on the CCD detectors ($D_1 - D_2$). Right: Simulated spectra calculated for a Sun--spacecraft distance of 1~a.u. using the {\sc chianti} database - top panel is the \fexxv\ spectrum with \fexxiv\ dielectronic satellites, lower panel is the \caxix\ spectrum with relatively much weaker \caxviii\ satellites. Temperatures (in MK) are indicated in the figure legend. }\label{chemix_simulated_spectra}
\end{figure}

Other subsystems of {\em ChemiX} include an energetic particle (electrons, protons, and alpha-particles) detector, so that the instrument will have its own monitor for safety concerns with the electronics; a soft X-ray pin-hole camera with CCD detector, providing X-ray context images with one-arcmin spatial resolution helping to ascertain which active region a flare originated from; and a movable target-pointing platform that can lock the spectrometer and CCDs on to a particular flare or other region within seconds of command from a flare trigger. Spectral atlases consisting of full spectral scans over the entire $1.5-9$\AA\ range will be taken while the instrument is directed at particular sources with a user-chosen intensity level (e.g. a total of $10\,000$ photon counts over the spectrum). Spectral line and continuum fluxes will be obtained with unprecedented precision, especially when the spacecraft is at perihelion, about 0.3 a.u. from the Sun. Thermal filters over the instrument entrance aperture will prevent solar ultraviolet radiation from entering.

\section{Conclusions}

The {\em DIOGENESS} instrument on {\em CORONAS-F} operated for only a few weeks in 2001 but several flares with {\em GOES} importance of up to X5 were recorded. Its scanning crystal spectrometers observed the neighbourhood of the He-like Ca, S, and Si (\caxix, \sxv, and \sixiii) X-ray emission lines at 3.17~\AA, 5.04~\AA, and 6.65~\AA, with the two quartz crystals of channels~1 and 4 in a dopplerometer mode, i.e. arranged in such a way that the Doppler line shifts at flare impulsive stages could be separated from spatial shifts. The concept was validated for the X5 flare on 25~August 2001, in which velocities of up to 150~km~s$^{-1}$ were observed. The strong X-ray emission during this flare enabled many lines to be distinguished and identified. The \caxix\ lines predominate in channels~1 and 4 spectra, with dielectronic satellites of \caxviii,  \caxvii, and \caxvi\ also present. The spectra are similar to those recorded with the X-ray spectrometer on the {\em P78-1} spacecraft \citep{dos85,see85}, and resemble to a high degree those obtained from the Alcator~C-Mod tokamak \citep{rice14}, including the Rydberg series of \arxvii\ $1s^2 - 1snp$ lines which are seen up to $n=7$ in {\em DIOGENESS} spectra. In channel~3 and 4 spectra, lines of \sxv, \sixiii, and \sixiv\ predominate, with dielectronic satellites of the Li-like stages occurring.

The {\em ChemiX} instrument on the forthcoming Russian {\em Interhelioprobe} spacecraft, which will be launched into elliptical orbits around the Sun in 2020 and 2022, will have four channels viewing \fexxv, \caxix, \sxv, \sixiii, and \sixiv\ X-ray emission lines, with a full coverage of the $1.5-9$~\AA\ range. The diffracting crystals will be arranged in a dopplerometer manner as for {\em DIOGENESS} but will be bent and non-moving, not flat and scanning. This will be a considerable improvement in that the time resolution should be much higher. Also the use of cooled CCD detectors will give much better spectral resolution over the {\em RESIK} spectrometer on {\em CORONAS-F} which had position-sensitive proportional counters. A collimator will allow the emission from individual active regions to be observed.

\begin{acks}
We acknowledge support from the Polish National Science Centre grants: 2011/01/B/ST9/05861,  4~T12E 045 29,  2011/01/M/ST9/06096, 2314/7.PR/2012/2 and the European Commission grant FP7/2007-2013: eHEROES, Project No.~284461
We also thank Rutherford Appleton Laboratory (UK) for the provision of the beryl crystal of {\em DIOGENESS}, which was a spare crystal from the construction of the {\em Solar Maximum Mission} Flat Crystal Spectrometer. The {\sc chianti} atomic database and code is a collaborative project involving George Mason University, University of Michigan (USA) and University of Cambridge (UK).

\end{acks}

%
\bibliographystyle{spr-mp-sola}
\bibliography{RESIK}

\end{article}
\end{document}